\documentclass{aa}  

\usepackage{graphicx}
\usepackage{afterpage}
\usepackage{float}
\usepackage{txfonts}
\usepackage{mhchem}
\usepackage{xcolor}

\newcommand{\kjmol}{kJ mol$^{-1}$\,}

\newcommand{\myhlight}[1]{#1}
\usepackage{hyperref}
\usepackage{natbib}
\bibpunct{(}{)}{;}{a}{}{,} 

\titlerunning{Interstellar Chemistry of CN Radicals on Ices: The formation of \ce{CH3CN} and \ce{CH3NC} and potential connection to acetamide}
\authorrunning{Enrique-Romero, J.; Lamberts, T.}

\begin{document} 

\title{Interstellar Chemistry of CN Radicals on Ices: \\ The formation of \ce{CH3CN} and \ce{CH3NC} and potential connection to acetamide}
\titlerunning{Interstellar Chemistry of CN Radicals on Ices: Formation of \ce{CH3CN} and \ce{CH3NC}}


   \author{J. Enrique-Romero*
          \inst{1}
          \, T. Lamberts \inst{1,2}
          }

   \institute{%
Leiden Institute of Chemistry, Gorlaeus Laboratories, Leiden University, PO Box 9502, 2300 RA Leiden, The Netherlands\\
\email{j.enrique.romero@lic.leidenuniv.nl}
\and
Leiden Observatory, Leiden University, P.O. Box 9513, 2300 RA Leiden, The Netherlands
}

   \date{}

\abstract
   {Among the most significant chemical functional groups of interstellar molecules are the class of nitriles, which are proposed as key prebiotic molecules due to their chemical connection to the peptide bond after hydrolysis. \ce{^.CN} radicals, the simplest representative of this group, have been shown to exhibit strong interactions with interstellar water ices, potentially impacting their reactivity with other radicals nearby.}
   {This study explores (a) whether CN and \ce{^.CH3} radicals can readily react to form methyl cyanide (\ce{CH3CN}) and its isomer methyl isocyanide (\ce{CH3NC}); and (b) the feasibility of the reaction \ce{(CN\cdots H_2O)_{hemi} \to ^{.}C(OH)=NH} and its potential role in the formation of acetamide.}
   {Following a benchmark, density functional theory was employed to map the potential energy surfaces of these chemical processes, focusing on their reactivity on water and carbon monoxide ices.}
   {The results show that CN reacts with \ce{^.CH3} radicals on water ices, forming \ce{CH3CN} and \ce{CH3NC} efficiently. However, these reactions are driven by diffusion of \ce{^.CH3} towards the reactive site and subsequently compete with back-diffusion of \ce{^.CH3} from that site. The formation of the radical intermediate \ce{^{.}C(OH)=NH} on water ice requires quantum tunnelling and assuming that acetimidic acid forms via \ce{CH3 + ^{.}C(OH)=NH \to CH3C(OH)=NH}, it can also only isomerize into acetamide through a sizable barrier thanks to quantum tunnelling. Both quantum tunnelling-driven reactions are highly dependent on the local structure of the water ice. Finally, radical coupling reactions on carbon monoxide ices are found to be barrierless for all cases and again, both the cyanide and the isocyanide are formed.}
  {This work reinforces the conclusion that \ce{^.CN} radicals on interstellar grain surfaces are highly reactive and unlikely to persist unaltered.}

   \maketitle
%

\section{Introduction}\label{sec:intro}

The origins of life in the Universe have likely been seeded in the cold and harsh conditions of the interstellar medium (ISM), where chemical evolution occurs alongside star and planet formation \cite{Ceccarelli2023ASPC}. With over 300 molecular species detected in the ISM, those containing a \ce{-CN} group have been proposed as important prebiotic parent molecules. This is due to the ability of the \ce{-CN} group to undergo hydrolysation reactions, becoming peptide bonds and forming amino acids \citep[][]{Goldman2010_CNAminoacids} or can be hydrogenated to become amines \citep{raaphorst_towards_2025}. Nitrile-bearing molecules comprise a large portion of interstellar molecule detections ($\sim$15\%), and in particular, simple cyanides such as hydrogen cyanide (HCN) or methyl cyanide \ce{CH3CN} (also called acetonitrile) are widely observed along different stages of the star and planet-forming processes.

This study focuses on the formation of methyl cyanide (\ce{CH3CN}) and its isomer, methyl isocyanide (\ce{CH3NC}). Methyl cyanide has been detected in the gas phase in a wide range of interstellar environments, including starless and prestellar cores \citep[e.g.,][]{Mejias2023MNRAS, Scibelli2024MNRAS}, low and high-mass protostars \citep[e.g.][]{Bergner2017ApJ, Nazari2021, Nazari2022, Bianchi2022}, protoplanetary disks \citep[e.g.,][]{Oberg2015Natur, Ilee2021ApJS} and Solar System bodies such as comets \citep[e.g.,][]{LeRoy2015_comets, Altwegg2019ARA&A}. In addition, methyl cyanide has recently been tentatively detected in interstellar ices \citep[][]{Nazari2024}.

This interstellar complex organic molecule (iCOM) can form either in the gas phase or on interstellar ices.

\myhlight{Recently, \citet{Giani2023MNRAS} and \citet{Mancini2024A&A} have revised the main gas-phase formation and evolution pathways of acetonitrile in the gas phase. Giani et al. identified ten reactions contributing to its synthesis, four of which are particularly significant: 
\ce{CH3^+ + HCN \to CH3CNH^+ + h\nu} and 
\ce{CH3OH2^+ + HNC \to CH3CNH^+ + H2O}, 
followed by either dissociative electron recombination or an acid-base reaction with \ce{NH3} to convert \ce{CH3CNH^+} into neutral \ce{CH3CN}.}

\myhlight{In contrast, the dominant non-energetic solid-phase formation mechanisms on ices are thought to involve the radical-radical coupling reactions} 
\ref{chem_eq:global_ch3+cn}a and \ref{chem_eq:global_ch3+cn}b
\citep{Garrod2006, Garrod2008}, 
\begin{subequations} \label{chem_eq:global_ch3+cn}
\begin{align}
    \text{\ce{CH3 + CN} } & \text{\ce{\rightarrow CH3CN},}
    \label{chem_eq:ch3+cn}\\
    \text{\ce{CH3 + CN} } & \text{\ce{\rightarrow CH3NC},}
    \label{chem_eq:ch3+cn_iso}
\end{align}
\end{subequations}
and the successive hydrogenation of \ce{CCN}, \citep[][reaction \ref{chem_eq:ccn+h}]{Loomis2018ApJ}:
\begin{align}
    \text{\ce{CCN + H} } & \text{\ce{\rightarrow \cdots \rightarrow CH3CN}.}
    \label{chem_eq:ccn+h}
\end{align}

Interestingly, observational studies of Class 0/I protostars in the Perseus molecular cloud have revealed a correlation between the column densities of methanol and methyl cyanide \citep{Yang2021ApJ}. This correlation may point to a shared origin from dust grain surface chemistry or to a chemical connection in the gas phase \citep{Giani2023MNRAS, Bianchi2022}.

\noindent We focus on reactions \ref{chem_eq:ch3+cn} and \ref{chem_eq:ch3+cn_iso}, building on the findings of CN hydrogenation by \citet{enrique2024complex}. That work showed CN hydrogenation leads to HCN and HNC efficiently on \ce{H2O} and CO ices. On water ice, CN interacts strongly through hemibonding, which nonetheless only minimally hinders reaction with atomic H, yet imposes a significant activation barrier for reactions with \ce{H2}. Hemibonding further facilitates the formation of \ce{HO^.CNH}, a precursor to formamide \citep{Rimola2018}. On CO ice, CN interactions are weaker, allowing barrierless hydrogenation with atomic H, while reactions with \ce{H2} retain a similar barrier ($\sim$12 kJ mol$^{-1}$) as in the gas phase. \ce{^.CN} can react with CO molecules too, forming \ce{NC^.CO}, which can barrierlessly get hydrogenated into \ce{HCOCN} by atomic hydrogen.

The goals of this work are to go beyond reactivity with hydrogen and thus investigate (a) how \ce{^.CN} radicals react with \ce{^.CH3} radicals on interstellar ices to form \ce{CH3CN} and \ce{CH3NC}, (b) the role played by diffusion of \ce{^.CH3} (explicitly looked into for the first time), and (c) the reactivity of \ce{^.CN} radicals with water molecules from the ice. We use computational quantum chemical tools, outlined in the methods section (Sect. \ref{sec:methods}), and the results are presented in Sect. \ref{sec:results}. In Sect. \ref{sec:discussion}, we examine the implications of our findings in the context of astrochemistry. And finally, Sect. \ref{sec:conclusions} summarises the main conclusions of this work.

\section{Methodology}\label{sec:methods}

Density functional theory (DFT) calculations were performed using ORCA 6.0.1 \citep{neese2020orca,Neese2022}. We have performed a benchmark study where different reference methods were employed. The explicitly correlated coupled-cluster CCSD(T)-F12 method \citep{knizia2009simplified_CCSDtF12} was used for non-radical-radical reactions and molecule-surface interactions, while the $N$-electron valence state perturbation theory (NEVPT2) \citep{angeli2001_nevpt2,angeli2007new_nevopt2} method was employed for radical-radical reactions, both using Molpro \citep{molpro1,molpro2}.
In addition, the complete active space perturbation theory up to second order (CASPT2) method \citep{werner1996_caspt2,celani2000_caspt2}, based on the automated construction of atomic valence active space (AVAS) \citep{sayfutyarova2017_AVAS}, was also used using both Molpro and OpenMolcas \citep{li2023openmolcas,battaglia2023multiconfigurational} to double-check our reference values for radical-radical reactions. All calculations employed triple-zeta quality basis sets including diffuse and polarisation functions (see below). For further information on the active space orbitals and raw energetics of the benchmark study, we refer to section Sect.\ref{sisec:benchmark} in the appendix.

\begin{table}[!htbp]
\centering
\caption{Benchmark on interactions and reactivity relevant for this work.}
\label{tab:benchmark_data}
\resizebox{0.9\columnwidth}{!}{%
\begin{tabular}{|c|cc|cc|}
\hline
Interactions        & \multicolumn{2}{|c|}{\ce{^.CH3-CO}} & \multicolumn{2}{c|}{\ce{^.CH3-(H2O)3}} \\
         & Case 1 & Case 2 & Case 1 & Case2 \\ \hline
{Method} &$E_b$ & $E_b$ & $E_b$ & $E_b$ \\ \hline
{B3LYP-D4} & -6.0 & -3.8 & -9.0 & -5.7 \\
{BHLYP-D4} & -10.9 & -9.2 & -7.9 & -5.3 \\
{M062X-D3} & -11.6 & -7.4 & -11.5 & -14.0 \\
{MPWB1K-D3BJ} & -10.1 & -8.1 & -7.7 & -6.5 \\
{M08HX-D3} & -9.1 & -6.4 & -8.6 & -10.0 \\
{M05X-D3} & -11.0 & -7.5 & -11.7 & -11.5 \\
{PW6B95-D4} & -8.2 & -6.0 & -9.0 & -8.1 \\
{RSCAN-D3BJ} & -6.4 & -3.2 & -9.5 & -8.6 \\
{wB97M-D3BJ} & -6.4 & -4.1 & -8.6 & -7.9 \\
{wB97M-V} & -6.1 & -4.1 & -8.1 & -6.9 \\
{wB97X-D4} & -5.9 & -3.7 & -8.3 & -6.8 \\ \hline
{CCSDt-F12/AVZ} & -3.1 & -1.8 & -8.0 & -5.9 \\ \hline
\end{tabular}%
}
\par\vspace{0.5cm}
\resizebox{\columnwidth}{!}{%
\begin{tabular}{|c|c|c|c|}
\hline
 Formation& \ce{CH3CN} & \ce{CH3NC} & \ce{^.CN + H2O} \\
 of:& on a \ce{(H2O)3} & on a \ce{(H2O)3} & \ce{\to HO^.CNH} \\ \hline
Method & $\Delta E_a$ & $\Delta E_a$ & $\Delta E_a$ \\ \hline
B3LYP-D4 & 3.1 & 3.7 & 35.3 \\
BHLYP-D4 & 3.5 & 17.6 & 41.5 \\
M062X-D3 & 1.9 & 13.1 & 26.5 \\
MPWB1K-D3BJ & 3.6 & 12.9 & 33.8 \\
M08HX-D3 & 3.2 & 12.4 & 39.1 \\
M05X-D3 & 3.7 & 77.2 & 23.8 \\
PW6B95-D4 & 3.3 & 45.7 & 37.1 \\
RSCAN-D3BJ & 3.4 & 3.8 & 19.8 \\
wB97M-D3BJ & 3.4 & 10.0 & 38.3 \\
wB97M-V & 3.4 & 10.4 & 34.7 \\
wB97X-D4 & 3.4 & 14.0 & 35.5 \\ \hline
CCSDt-F12/AVZ & -- & -- & 43.8 \\
CASPT2/AVZ* & 3.5 & 4.1 & -- \\
CASPT2/AVZ$^\dagger$ & 3.4 & 2.8 & -- \\
NEVPT2/AVZ$^\dagger$ & 3.6 & 2.7 & -- \\ \hline
\end{tabular}%
}
\tablefoot{
wHt and AVZ stand for water-assisted H-transfer and AUG-CC-PVTZ, respectively. Energy units in \kjmol. The def2-PVTZD basis set was used for all DFT-D calculations. More data can be found in Sect. \ref{sisec:benchmark} of the appendix.
\tablefoottext{*}{{\sc OpenMolcas} CASPT2(15e,24o)/AVTZ (see appendix).}
\tablefoottext{${\dagger}$}{{\sc Molpro} CASPT2/NEVTP2(12e,10o)/AVTZ (see appendix).}
}
\end{table}

Our benchmark concludes that the BHandHLYP functional \citep{Becke1993}, combined with the D4 correction \citep{caldeweyher2017extension} for long-range interactions and the def2-TZVPD basis set \citep{hellweg2015development}, provides an appropriate level of theory for modelling both the \ce{^.CH3 + ^.CN} reactivity and \ce{^.CH3}--\ce{H2O} interactions.
There are two cases where BHandHLYP underperforms: the interaction of \ce{^.CH3} with carbon monoxide \citep[as previously reported by][]{Lamberts2019} and the \ce{^.CH3 + ^.CN \to CH3NC} reaction. This is not a problem for the former as will be discussed in Sect.\ref{ssec:COreactivity}. For the latter, the meta-GGA RSCAN-D3(BJ) functional shows better performance and therefore will be used for this particular reaction. 

All DFT calculations encompass geometry optimisations, transition state searches, intrinsic reaction coordinate (IRC) calculations to connect stationary points along the reaction pathway on the potential energy surface (PES), PES scans, nudged elastic band (NEB) calculations \citep{asgeirsson2021nudged}, and frequency calculations to ensure that our geometries correspond to minima or first-order saddle points for transition states. Tight self-consistent field (TightSCF) convergence criteria and dense integration grids (DEFGRID3) were adopted throughout this work in {\sc Orca}.
Unrestricted Kohn-Sham orbitals were used consistently, and for all open-shell singlet calculations, we applied the broken spin symmetry approach using {\sc Orca}’s spin-flip routine.

\section{Results}\label{sec:results}

In this section, we break down all the results obtained for the reactivity of \ce{^.CN} with \ce{^.CH3} on \ce{H2O} and CO ices, as well as the reactivity of \ce{^.CN} with a water molecule leading to the intermediate radical \ce{HO^.CNH} which may subsequently react with \ce{^.CH3}. A summary of all the reactions studied can be found in Table \ref{tab:energies}.

\begin{table*}[!htbp]
\centering
\caption{Summary of the reactions studied in this work, organised according to the binding mode of \ce{^.CN} on either water or carbon monoxide.}
\label{tab:energies}
\resizebox{0.9\textwidth}{!}{%
\begin{tabular}{llcccc}
\multicolumn{6}{c}{Hemibonded \ce{^.CN} on \ce{H2O}} \\ \hline
Reaction     & Product & Barrier & Reaction energy & Trans. freq. & Crossover temp.\\ \hline
\ce{^.CH3 + ^.CN_{hemi}} & \ce{CH3CN}   & 2.8     & -495.1 & 117 & --\\
\ce{^.CH3 + ^.CN_{hemi}} & \ce{CH3NC}   & 3.1$^{\delta}$    & -355.7$^{\delta}$ & 100 & --\\
\ce{^.CN_{hemi} + 2H2O}& \ce{t-HO^.CNH}$_{\text{wHt(2)}}$    & 51.6 & -120.1 & 592 & 135\\
\ce{^.CN_{hemi} + 3H2O} & \ce{c-HO^.CNH}$_{\text{wHt(3)}}$ & 34.1 & -84.6 & 310 & 71 \\
\ce{^.CN_{hemi} + 4H2O} & \ce{t-HO^.CNH}$_{\text{wHt(4)}}$ & 20.7 & -109.4 & 550 & 126\\ 
\ce{^.CH3 + t-HO^.CNH}$_{\text{wHt(2)}}$ & \ce{CH3COHNH} & 2.1$^{\alpha}$ & -372.8$^{\alpha}$ & -- & -- \\
\ce{^.CH3 + c-HO^.CNH}$_{\text{wHt(3)}}$ & \ce{CH3COHNH} & 1.2$^{\alpha}$ & -389.8$^{\alpha}$ & -- & -- \\
\ce{^.CH3 + t-HO^.CNH}$_{\text{wHt(4)}}$ & \ce{CH3COHNH} & 3.1$^{\beta}$ & -351.8$^{\beta}$ & -- & -- \\
\hline
 &  &  &  \\
\multicolumn{6}{c}{\ce{^.CH3} diffusion (averaged values)} \\ \hline
Reaction     & Product & Barrier & Reaction energy & Trans. freq. & Crossover temp.\\ \hline
Diffusion \ce{(^.CH3)}$_{(H_2O)_{14} + \cdot CN}$ & -- & 2.3 & 1.0 & -- & --\\ 
Diffusion \ce{(^.CH3)}$_{(H_2O)_{14}}$  & -- & 3.2  & 0.2 & -- & --\\  \hline
 &  &  &  \\
\multicolumn{6}{c}{Water-catalysed Acetimidic acid--Acetamide conversion} \\ \hline
Reaction     & Product & Barrier & Reaction energy & Trans. freq. & Crossover temp.\\ \hline
AAc $\to$ Am (BS$_1$) & \ce{CH3C(O)NH2} & 58.1 & -44.9 & 255 & 58\\ 
AAc $\to$ Am (BS$_2$) & \ce{CH3C(O)NH2} & 20.0 & -61.2 & 584 & 134\\
\hline
 &  &  &  \\
\multicolumn{6}{c}{Formation of cyanic (\ce{HOCN}) and isocyanic acid (\ce{HNCO}) on water ices} \\ \hline
Reaction     & Product & Barrier & Reaction energy & Trans. freq. & Crossover temp.\\ \hline
t-\ce{HO^.CNH}$_{\text{wHt(4)}}^{\gamma}$ + \ce{^.CH3} & \ce{HOCN + CH4} & 23.3 & -303.1 & 2029 & 464 \\ 
t-\ce{HO^.CNH}$_{\text{wHt(4)}}$ + \ce{^.CH3} & \ce{HNCO + CH4} & 33.2 & -384.9 & 3237 & 741 \\ 
t-\ce{HO^.CNH}$_{\text{wHt(4)}}$ + \ce{^.H} & \ce{HOCN + H2} & 6.0 & -310.7 & 1504 & 344 \\ 
t-\ce{HO^.CNH}$_{\text{wHt(4)}}$ + \ce{^.H} & \ce{HNCO + H2} & 23.9 & -398.9 & 3204 & 734 \\ \hline
 &  &  &  \\ 
\multicolumn{6}{c}{H-bonded \ce{^.CN} on \ce{H2O}} \\ \hline
Reaction     & Product & Barrier & Reaction energy & Trans. freq. & Crossover temp.\\ \hline
\ce{^.CH3 + ^.CN}$^{\varepsilon}$ & \ce{CH3CN} & Barrierless & -532.9 & -- & --\\
\ce{^.CH3 + ^.CN} & \ce{CH3NC} & No reaction & -- & -- & --\\ \hline
 &  &  &  \\ 
\multicolumn{6}{c}{\ce{^.CN} van der Waals bonded on CO} \\ \hline
Reaction     & Product & Barrier & Reaction energy & Trans. freq. & Crossover temp.\\ \hline
\ce{CH3 + CN}$^{\varepsilon}$ & \ce{CH3CN} & Barrierless & -524.0 & -- & --\\ 
 \ce{CH3 + CN}$^{\varepsilon}$ & \ce{CH3NC} & Barrierless & -422.4 & -- & --\\ \hline
 &  &  &  \\
\multicolumn{6}{c}{Covalently bound \ce{NC^.CO}} \\ \hline
Reaction     & Product & Barrier & Reaction energy & Trans. freq. & Crossover temp.\\ \hline
\ce{^.CH3 + NC^.CO} & \ce{CH3C(O)CN} & Barrierless & -493.4 & -- & --\\ \hline
\end{tabular}%
}
\tablefoot{
Energy units are in kJ mol$^{-1}$. wHt stands for ``water-assisted H-transfer''. AAc stands for acetimidic acid (\ce{CH3COHNH}), while Am for acetamide (\ce{CH3CONH2}). Raw energetics and extra data can be found in Sect.\ref{sisec:energetics_onice} of the appendix.
\tablefoottext{${\alpha}$}{Energy relative to the intermediate (i.e., the product of the first reaction step, wHt(2,3)).}
\tablefoottext{${\beta}$}{Energy relative to a manually built intermediate with an energy relative to the products (wHt(4)) of the first step of -6.4 \kjmol.}
\tablefoottext{${\gamma}$}{Reactants structure prepared manually, see Figure \ref{fig:cyanic_acid}.}
\tablefoottext{${\varepsilon}$}{Reactants' geometry was prepared putting \ce{^.CH3} far from \ce{^.CN}.}
\tablefoottext{${\delta}$}{Energetics calculated at RSCAN-D3(BJ) level, see Sect.\ref{sec:methods}.}
}
\end{table*}

\subsection{Reactivity on water ices}

\paragraph{Radical-radical coupling:} 
The principal binding mode of \ce{^.CN} radicals in water is through the formation of hemibonded complexes, that is, two-centre and three-electron bonds \citep{enrique2024complex}. Based on the high reactivity of \ce{^.CN} in this binding mode with H atoms \citep{enrique2024complex}, here, we consider the possibility that it may react with a \ce{^.CH3} radical instead. In this case, we consider the reactions of \ce{C-C} and \ce{C-N} bond formation. As shown in Table \ref{tab:energies}, both channels sport an activation energy barrier. The mechanism involves breaking the \ce{(H2O-^.CN)_{hemi}} hemibond as the \ce{C-C} or \ce{C-N} bonds form. The barriers for \ce{C-C} and \ce{C-N} bond formation are 2.8 \kjmol and 3.1 \kjmol, respectively (notice the latter was calculated at RSCAN-D3(BJ) level, see Sect.\ref{sec:methods} and Table \ref{tab:benchmark_data}).\\

\begin{figure*}[!htb]
    \centering
    \includegraphics[width=0.7\textwidth]{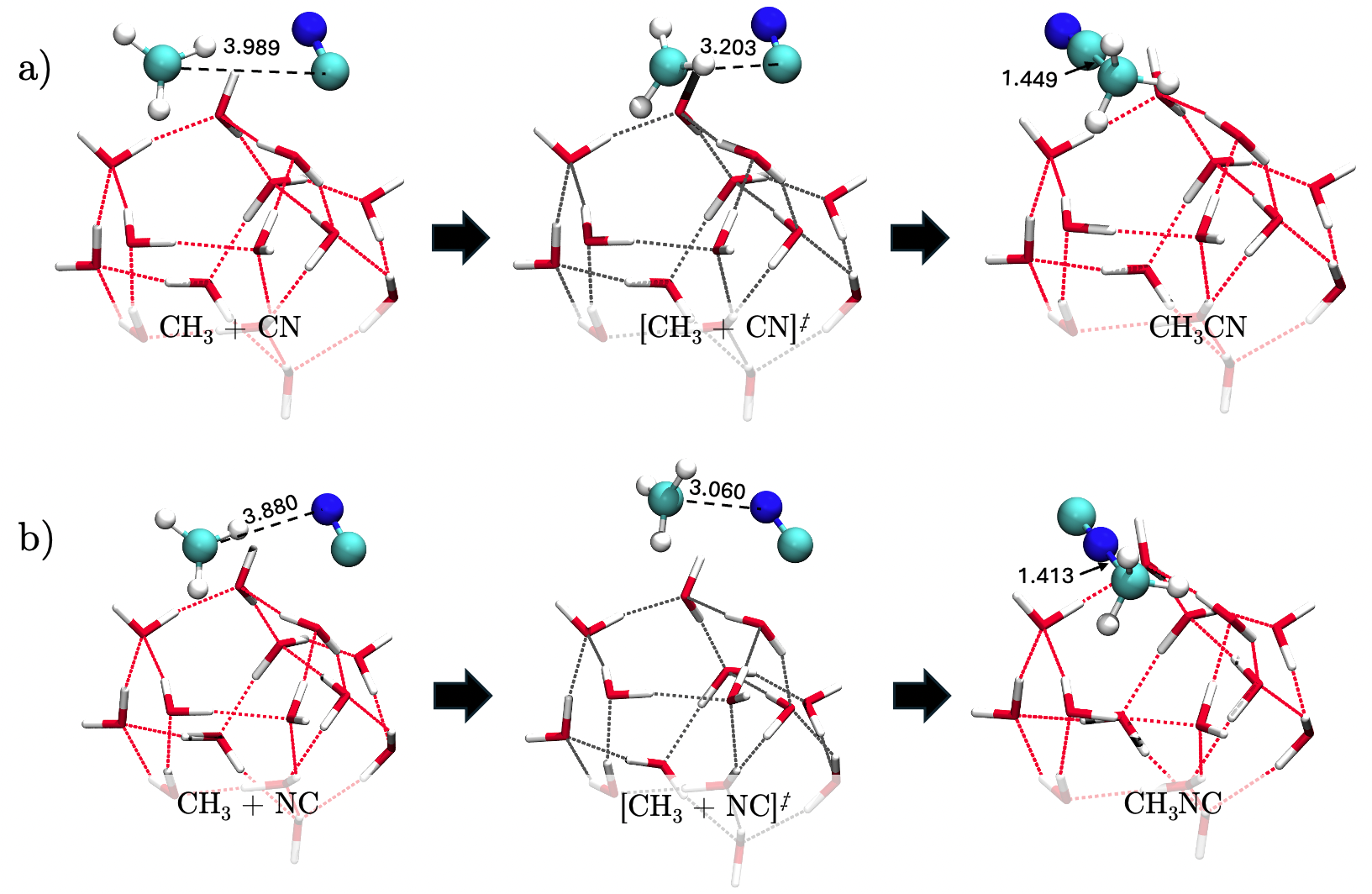}
    \caption{Snapshots of the potential energy surface stationary points for the formation of (a) \ce{CH3CN} and (b) \ce{CH3NC} on the \ce{(H2O)14} ice cluster, where the \ce{^.CN} radical is initially hemibonded to the ice. Distances in \AA. Atoms relevant to the reaction are highlighted using a ball-and-stick representation. Atom colour code is: red for O, blue for N, grey for C and white for H.}
    \label{fig:ch3cn_ch3nc}
\end{figure*}

In addition to the hemibond, the binding energy distribution of \ce{^.CN} also includes a weaker, hydrogen-bonded mode. However, this mode has much less astrochemical significance, as it constitutes only a minimal portion of the binding energy distribution. This interaction leads to a barrierless coupling, consistently proceeding through the \ce{C-C} bond-forming channel toward \ce{CH3CN}. Notably, the absence of a barrier implies that the rate-determining step of the reaction is the diffusion of \ce{^.CH3}.

\begin{figure*}[!htbp]
    \centering
    \includegraphics[width=0.7\textwidth]{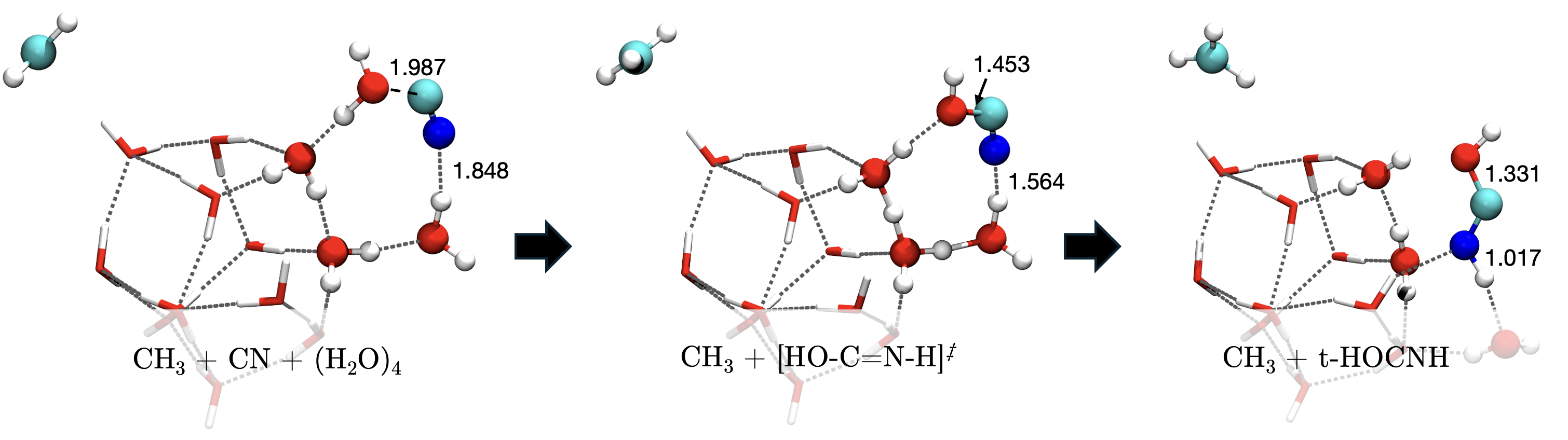}\\
    {\small \textbf{(a)} First step, water-assisted H-transfer (wHt(4))}\\
    \par\vspace{0.2cm}
    \includegraphics[width=0.7\textwidth]{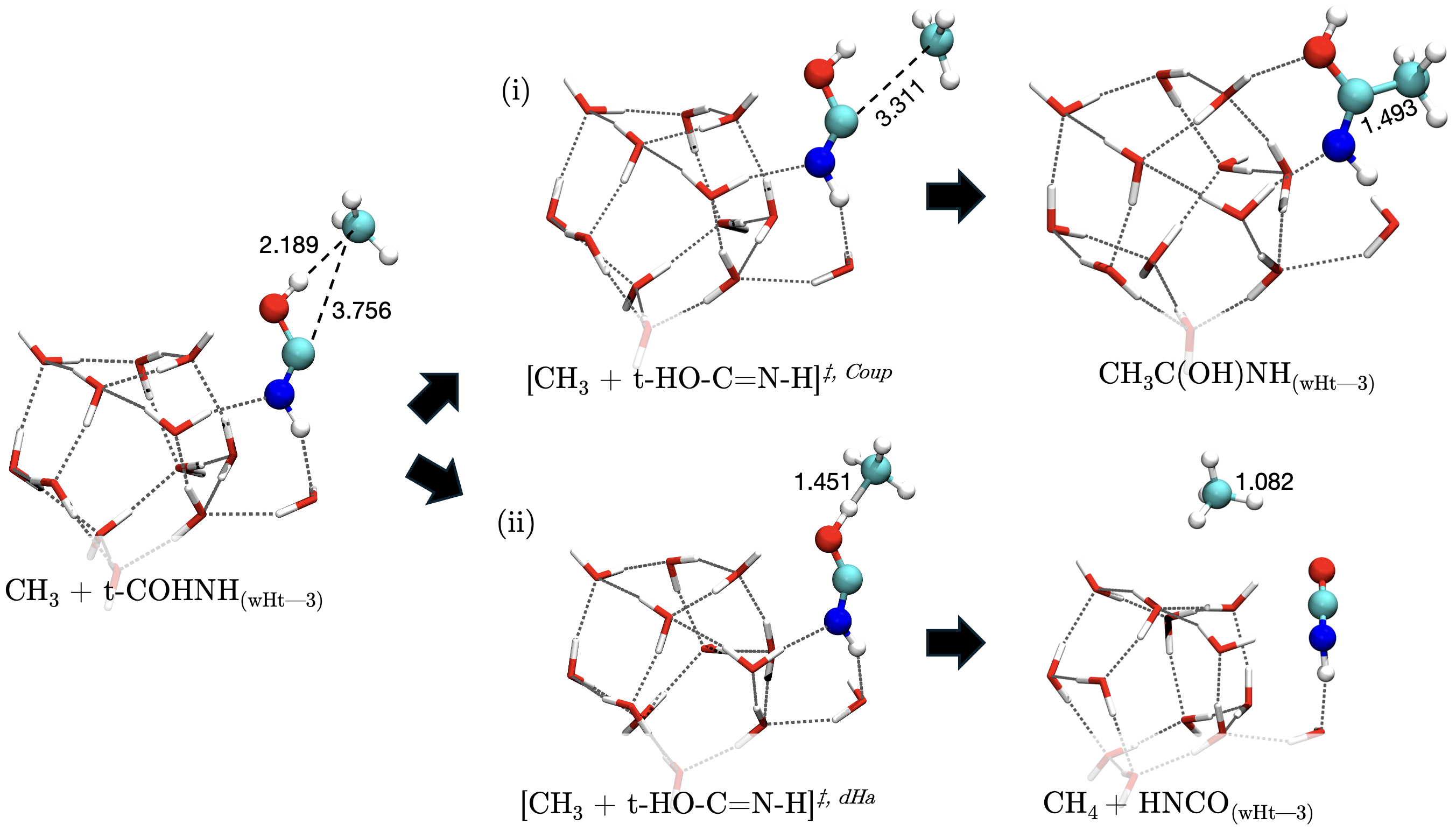}\\
    {\small \textbf{(b)} Second step: i) coupling with \ce{^.CH3} to give acetimidic acid (AAc)}\\
    {\small and ii) direct H-transfer to form \ce{CH4 + HNCO}}\\
    \caption{Snapshots of the potential energy surface stationary points for (a) the \ce{(H2O-^.CN)_{hemi} \to HO^.C=NH} reaction (wHt(4)), and (b) a second step where acetimidic acid is formed via \ce{^.CH3 + HO^.C=NH \to CH3C(OH)NH} (i) and an alternative path leading to \ce{CH4 + HNCO} through a direct H-transfer reaction. Distances are provided in \AA. Atoms relevant to the reaction are highlighted using a ball-and-stick representation. Atom colour code is: red for O, blue for N, grey for C and white for H.}
    \label{fig:wht_4water}
\end{figure*}

\paragraph{Water-assisted COM formation}
Besides radical couplings, an alternative reaction pathway involves direct reaction with the water matrix. In this case, the hemibonded complex \ce{(H2O-^.CN)_{hemi}} can evolve into the formamide precursor \ce{HO-C^.=N-H} mediated by quantum tunnelling, following the mechanism proposed by \citeauthor{Rimola2018}, through water-assisted hydrogen transfer reactions. This scenario is at play if the hemibonded complex is sufficiently long-lived, e.g., it lies in the ice matrix or is not readily destroyed by competing chemical reactions. Depending on the length of the water-assisted H-transfer (wHt) chain, we find that the activation energy barrier for transforming \ce{(H2O-^.CN)_{hemi}} into \ce{HO-^.C=NH} varies.
For hydrogen-bonded \ce{H2O} chains consisting of 2, 3, and 4 water molecules—denoted as wHt(2), wHt(3), and wHt(4) (see Figure~\ref{fig:wht_4water} for wHt(4)), the calculated reaction barriers are 51.6, 34.1, and 20.7~\kjmol, respectively. These processes involve the concerted motion of multiple hydrogen atoms and exhibit relatively low but notable tunnelling crossover temperatures of 136, 71, and 126~K, respectively.

This mechanism potentially enables further reactions of the intermediate \ce{HO-C^.=NH}, such as with a nearby radical like \ce{^.CH3}:

\begin{align}
\text{\ce{CH3 + HO-^.C=NH}} &\rightarrow \text{\ce{HO-C(CH3)=NH}}, \label{chem_eq:ch3+hocnh} \\
\text{\ce{CH3 + HO-^.C=NH}} &\rightarrow \text{\ce{CH4 + HOCN}}, \label{chem_eq:ch3+hocnh__cyanic_acid} \\
\text{\ce{CH3 + HO-^.C=NH}} &\rightarrow \text{\ce{CH4 + OCNH}.} \label{chem_eq:ch3+hocnh__isocyanic_acid} 
\end{align}

Reaction~\ref{chem_eq:ch3+hocnh} leads to the formation of acetimidic acid (\ce{HO-C(CH3)=NH}), a low-barrier product and one of the \ce{C2H5NO} isomers, along with acetamide and 1-aminoethanol (see Section~\ref{sec:discussion}). 
We found the barriers for reaction~\ref{chem_eq:ch3+hocnh} proceeding from each wHt(i)-formed \ce{HO-C^.=NH} intermediate to be quite low: 2.1, 1.2, and 3.1~\kjmol. These are comparable to the barriers reported for reactions~\ref{chem_eq:ch3+cn} and~\ref{chem_eq:ch3+cn_iso}, and primarily involve the reorientation of the \ce{^.CH3} radical. In our simulations, the radical was initially positioned close to the reactive site, thus eliminating the need to consider local diffusion barriers toward the \ce{HO^.CNH} intermediate (see Figure~\ref{fig:wht_4water}).

\begin{figure}
    \centering
    \includegraphics[width=\columnwidth]{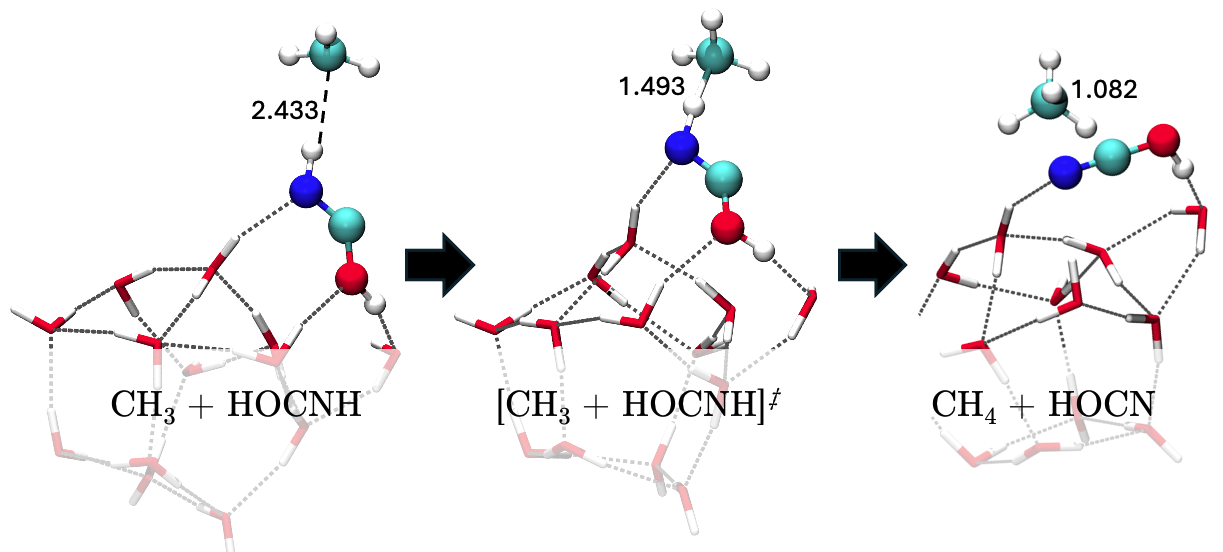}
    \caption{Snapshots of the potential energy surface stationary points to form cyanic acid (\ce{HOCN}) from the direct H-abstraction reaction \ce{HO^.CNH + CH3} on the \ce{(H2O)14} cluster. The reactants geometry was prepared manually to mimic that in \ref{fig:wht_4water}(b). Distances in \AA. Atoms relevant to the reaction are highlighted using a ball-and-stick representation. Atom colour code is: red for O, blue for N, grey for C and white for H.}
    \label{fig:cyanic_acid}
\end{figure}

On the other hand, reactions~\ref{chem_eq:ch3+hocnh__cyanic_acid} and~\ref{chem_eq:ch3+hocnh__isocyanic_acid} result in the formation of cyanic and isocyanic acids along with methane, respectively, through a direct H-transfer (dHa)., depicted in Figures~\ref{fig:wht_4water}(b-ii) and~\ref{fig:cyanic_acid}, respectively. These proceed over barriers of 33.2 and 23.3~\kjmol and are associated with high tunnelling crossover temperatures of 740 and 464~K, suggesting that quantum tunnelling may significantly enhance their reaction rates. Since none of the wHt(i) products displayed a conformation with the \ce{-NH} group exposed, the reactant geometry shown in Figure~\ref{fig:cyanic_acid} was constructed manually, mirroring the configuration in Figure~\ref{fig:wht_4water}(b).

Lastly, it is worth noting that in the work of \citet{Rimola2018}, a second wHt step was proposed, converting \ce{HO^.CNH} into \ce{O^.CNH2}. This transformation requires the \ce{-OH} group of \ce{HO^.CNH} to be hydrogen-bonded to the \ce{-NH} group via a bridging water network. Such a configuration is absent in our wHt(4) case, and longer water chains could result in higher barriers, potentially exceeding the 6.4~\kjmol barrier reported by Rimola and co-workers for their intermediate.

\subsection{Diffusion of \ce{^.CH3} on \ce{H2O}}

All coupling reactions described above share a common characteristic: they are mediated by \ce{^.CH3} diffusion towards the reactive site, and once there, there is a competition between the reaction and the diffusion of highly mobile \ce{^.CH3} away from the reaction site. To investigate this behavior, we calculated diffusion barriers for \ce{^.CH3} on the water ice clusters. We study the potential effect of having the cyanide radical hemibonded on the surface by testing two scenarios: one with \ce{^.CN} hemibonded to the ice surface (\ce{^.CN-(H2O)14}) and one on the bare \ce{(H2O)14} ice . For a more detailed look at the diffusion stationary points and energetics, we refer the reader to appendix \ref{sifig:SIdiff}.

\begin{figure}[!htbp]
    \centering
    \includegraphics[width=0.8\columnwidth]{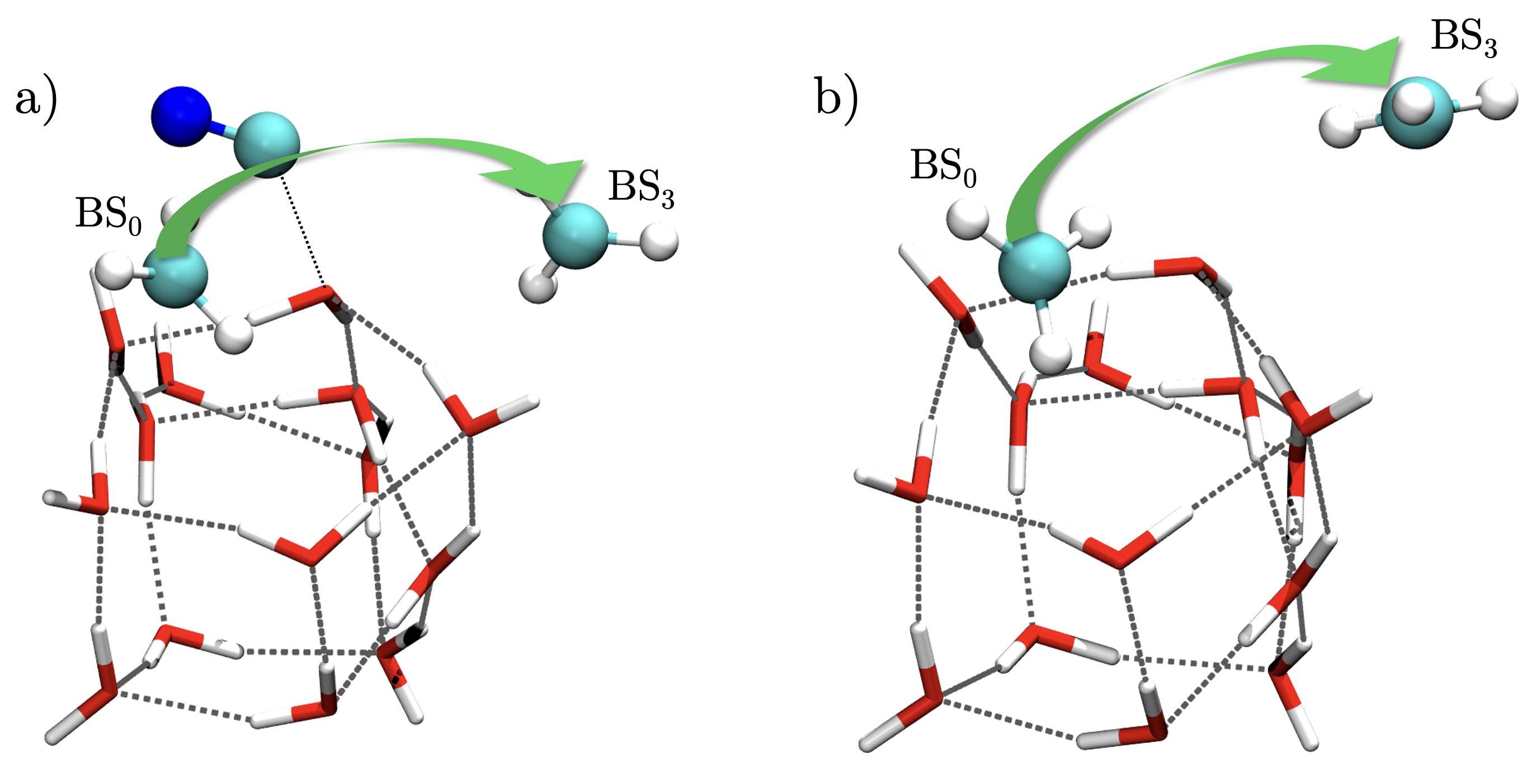}
    \caption{Diffusion of \ce{^.CH3} from the initial binding site (BS$_0$) to the neighbouring binding site BS$_3$ for two scenarios: a) \ce{^.CN-(H2O)14} and the bare ice b) \ce{(H2O)14}. The arrows indicate the motion the \ce{CH3} radical follows when it hops to BS$_3$. The initial BS$_0$ geometry in the left panel corresponds to the reactants geometry for the formation of \ce{CH3CN} and \ce{CH3NC} in Figure \ref{fig:ch3cn_ch3nc}. Atoms relevant to the reaction are highlighted using a ball-and-stick representation. Atom colour code is: red for O, blue for N, grey for C and white for H.}
    \label{fig:diff}
\end{figure}

We prepared the initial binding site taking the reactant structure for \ce{CH3CN} formation (Figure \ref{fig:ch3cn_ch3nc}a), with transition states identified for \ce{^.CH3} moving one binding site further from \ce{^.CN}. For the case without \ce{^.CN} on the ice surface, \ce{^.CH3} started from the same binding site, where the cyanide radical was removed and the ice reoptimised. See Table~\ref{tab:diffusion} and Figure~\ref{fig:diff} for the energies and examples of the diffusion trajectory. 
In both cases, we reproduce the displacement of \ce{^.CH3} to the nearest four neighbouring sites. 

\begin{table}[!htbp]
\centering
\caption{Diffusion energetics away from the same binding site for the case with \ce{^.CN} present on the ice, and without (see Figure \ref{fig:diff}).}
\label{tab:diffusion}
\resizebox{\columnwidth}{!}{%
\begin{tabular}{|c|cc|c|cc|}
\hline
 & \multicolumn{2}{c|}{\ce{^.CN-(H2O)14 + ^.CH3}} & & \multicolumn{2}{|c|}{\ce{(H2O)14 + ^.CH3}} \\ \hline
 & $\Delta E_{a}$ & $\Delta E_{r}$ & & $\Delta E_{a}$ & $\Delta E_{r}$ \\ \hline
BS$_1$ & 1.1 (0.9) & -0.1 &  BS$_1$ & 2.3 (2.3) & -2.0 \\
BS$_2$ & 3.0 (2.4) &  2.0 &  BS$_2$ & 0.7 (1.2) &  1.0 \\
BS$_3$ & 1.8 (2.1) &  1.7 &  BS$_3$ & 2.5 (4.0) &  0.1 \\
BS$_4$ & 3.1 (3.9) &  0.7 &  BS$_4$ & 4.1 (3.8) & -0.6 \\ \hline
\end{tabular}%
}
\tablefoot{
All energy values are in \kjmol and contain zero-point vibrational energy corrections except the ones in parentheses, which are the bare DFT-D values.
}
\end{table}

Comparing the BHandHLYP-D4 energies in Table \ref{tab:diffusion} should take into account that these diffusion pathways are more likely to take place in the exothermic direction. Therefore, while the average barrier height seems to be somewhat lower for moving away from an \ce{^.CN} site (2.3 \kjmol) and moving on a pure water cluster (2.8 \kjmol), most pathways leading away from the cyanide are endothermic. In other words, once a methyl radical is close to the cyanide reactant, it is less likely to leave and we expect back-diffusion to be effectively suppressed. 

Including ZPVE corrections has only a small effect, of at most 0.8 \kjmol. We do caution the reader that the transition states identified in these diffusion calculations are characterised by very low (harmonic) imaginary frequencies, $\sim$10–90 cm$^{-1}$, and should therefore be interpreted with care. Additionally, the ZPVE is likely overestimated because long-range Van der Waals interactions are typically anharmonic. Nevertheless, we have confirmed that for all cases, the imaginary frequency corresponds to a translational mode of \ce{^.CH3} parallel to the surface. 

Astrochemical models typically approximate diffusion barriers as a fixed fraction of a molecule’s binding energy, commonly using values between 0.3 and 0.5. For species adsorbed on water ices, despite there being no fundamental physical justification for such a uniform scaling factor choice \citep[][]{Cuppen2024_review}. In astrochemical databases, typical binding energies for the methyl radical are 9.8~\kjmol \citep[UMIST;][]{McElroy2013} and 13.3~\kjmol \citep[KIDA;][]{Kida2015}. On the other hand, computational studies provide broader estimates for the binding energy of \ce{^.CH3} on amorphous water ice: e.g., \citet{Ferrero2020} report values ranging from 9.1--13.3~\kjmol, while \citet{Duflot2021_binding} find a wider span of 7.7--16.1~\kjmol. Applying the standard 0.5 scaling \citep[e.g., as used in UCLCHEM;][]{Holdship2017_UCLchem} to the lowest reported binding energy (7.7~\kjmol) yields a diffusion barrier of approximately 3.8~\kjmol. This value is still higher than the diffusion barriers obtained in this work, suggesting that the commonly used approximation may underestimate the mobility of certain radicals under interstellar conditions.

\subsection{Reactivity on carbon monoxide ices}\label{ssec:COreactivity}
The reactivity on carbon monoxide ice is much simpler and similar to the situation without any explicit surface. On a \ce{(CO)13} cluster model \citep[][]{enrique2024complex}, the weakly bound \ce{^.CN} radical on the surface reacts with \ce{^.CH3} to form either \ce{CH3CN} and \ce{CH3NC} barrierlessly, depending on the relative orientation of the radicals upon encounter. This has been confirmed via relaxed PES scans starting where the \ce{C-C} and \ce{C-N} bonds are stretched up to 3.0 \AA. The only difference lies in the relative stability of each product, where \ce{CH3CN} is lower in energy than \ce{CH3NC}.

Additionally, as described in \citet{enrique2024complex}, \ce{^.CN} radicals may react with surface CO molecules forming \ce{NC^.CO} radicals with a small activation energy barrier of 2.5 \kjmol. If this intermediate form and is not destroyed by subsequent reactions such as H-addition, it can react with nearby radicals, such as \ce{^.CH3}. Indeed, we find that acetyl cyanide (\ce{CH3C(O)CN}) can be formed barrierlessly too.

The overestimation arising from the use of BHandHLYP-D4 for the\ce{^.CH3} interaction to carbon monoxide (see Sect.\ref{sec:methods}) does not impact the results, since a barrierless pathway is identified regardless of the interaction strength. In other words, despite the interaction being predicted to be too strong, this seems not to hamper reactivity. There, we decided not to recalculate the trajectories.

\section{Discussion and astrophysical implications}\label{sec:discussion}

\begin{figure}
    \centering
    \includegraphics[width=\columnwidth]{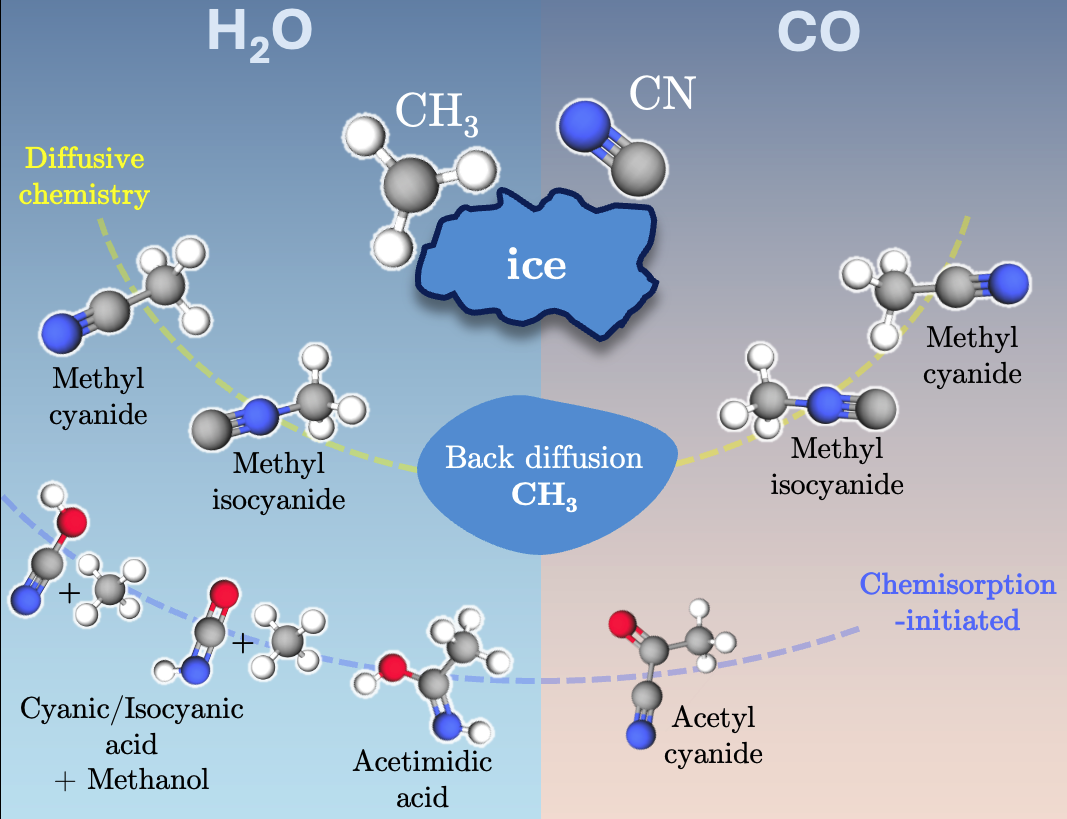}
    \caption{Summary of the reactions studied in this work between \ce{^.CH3 + ^.CN} on water and carbon monoxide ice surfaces. The first line indicates the possible outcomes of diffusive chemistry, while the second indicates the products after the reaction with one of the ice molecules from the surface. \myhlight{Notice in the chemisorption initiated products can proceed from two different mechanisms: chemisorption followed by H-abstraction (cyanic and isocyanic acid) or chemisorption followed by radical-radical coupling (acetimidic acid and acetyl cyanide).}}
    \label{fig:summary}
\end{figure}

Our results are summarised in Table \ref{tab:energies} and Figure \ref{fig:summary}. The most important result is that for \ce{^.CN} hemibonded to water (its most important binding mode), both the carbon and nitrogen atoms are available for reaction to form \ce{CH3CN} and \ce{CH3NC} with \ce{^.CH3}, respectively. Hence, the hemibond does not impose any chemical restrictions on radical coupling. This aligns with our previous findings for the reactions \ce{^.CN + ^.H \to HCN} and \ce{^.CN + ^.H \to HNC}. 

For this radical coupling mechanism, we adopt the common assumption in radical-radical chemistry regarding the availability of free H atoms on the ice. Specifically, this mechanism is expected to be active at intermediate temperatures, where the residence time of H atoms, the primary destroyers of radical species on the ice, is very short. Consequently, the formation of HCN and HNC would have occurred at lower temperatures, preceding that of \ce{CH3CN} and \ce{CH3NC}.

Even if the abundance of H atoms on the surface is low enough to allow the reactivity of radicals and \ce{CH3} is mobile, the activation energy for the formation of either \ce{CH3CN} and \ce{CH3NC} are as low as the diffusion barriers for \ce{CH3}. This is so because these barriers stem from the break of the weak interactions of mainly \ce{CH3} to the surface to move into the same binding site as \ce{CN}. Consequently, the competition between product formation and back diffusion plays a crucial role in these reactions, which could affect the final efficiency of these reaction paths. This of course also applies to the chemistry on CO ice, where the interactions of either \ce{CH3} and \ce{CN} are even weaker.
In any case, we note that the radical coupling reactions under discussion are not the only pathway for the formation of methyl cyanide and isocyanide. For example, on surfaces, reaction \ref{chem_eq:ccn+h} also plays a role, while additional pathways exist in the gas phase (see \ref{sec:intro}). Determining the relative importance of these pathways a priori is challenging, as it requires a dedicated study of reaction \ref{chem_eq:ccn+h} and astrochemical modelling. These two aspects will be the focus of a forthcoming work, and we defer this discussion to a later stage.

Regardless of its formation mechanism, methyl cyanide is a parent species for other molecules with high prebiotic interest in the ISM. One such molecule is acetamide (Am, \ce{CH3C(O)NH2})
, first detected nearly two decades ago in the star-forming region Sagittarius B2(N) \citep{hollis2006detection}, where it is abundant, comparable to formamide, with which it is consistently correlated, and acetaldehyde \citep{Halfen2011}. Notably, acetamide is one of the largest interstellar molecules detected with a peptide bond -- so far the largest is glycolamide by \citet{Rivilla2023ApJ}. More recently, acetamide has been observed in various protostellar environments \citep[e.g.,][]{Colzi2021_GUAPOS, Ligterink2022, Zeng2023_acetamide}, as well as on comet 67P/Churyumov–Gerasimenko by the Rosetta mission \citep{Goesmann2015, Altwegg2017}. Several formation pathways for acetamide have been proposed, including gas-phase ion-molecule reactions \citep{Quan2007, Halfen2011, Redondo2014}, ice-surface processing \citep[e.g.,][]{Frigge2018, Ligterink2018, Drabkin2023MNRAS}, and hydrogen abstraction from formamide (\ce{HCONH2}), followed by reaction with \ce{^.CH3} \citep{Belloche2017_acetamide}. 
We propose an alternative path to form acetamide. As shown by \citet{Rimola2018}, hemibonded \ce{^.CN} can evolve into \ce{OH-^.C=NH} via a water-assisted H-transfer mechanism. We have reproduced this mechanism, showing its dependence on the length of the H-transfer relay length of the water ice, which in turn also affects the isomery of the radical (cis or trans-\ce{HO^.CNH}). 
Depending on the water chain length and the method of choice, barriers ranging between 16.1 and 70.6 kJ/mol have been found, including also the works by \citet{Rimola2018} and \citet{Silva-Vera2024ESC}, and we argue that the lower barriers can be overcome thanks to quantum tunnelling. 
Once \ce{OH-^.C=NH} is formed, \citet{Rimola2018} proposes a follow-up water-assisted H-transfer reaction (with a local barrier of 6.4 \kjmol), which should again somewhat depend on the local properties of the ice structure. Instead, we propose that once \ce{OH-^.C=NH} is formed, it may react with a nearby radical, in this work a \ce{^.CH3} radical, forming acetimidic acid (AAc, \ce{CH3C(OH)NH}), a higher energy isomer of acetamide. For the three cases presented, the barrier of this follow-up reaction is of the same order as the barriers for the diffusion of \ce{CH3}, between 1.2--3.1 \kjmol (see Table \ref{tab:energies}, and the SI for more information).

\begin{figure}[!htbp]
    \centering
    \includegraphics[width=\columnwidth]{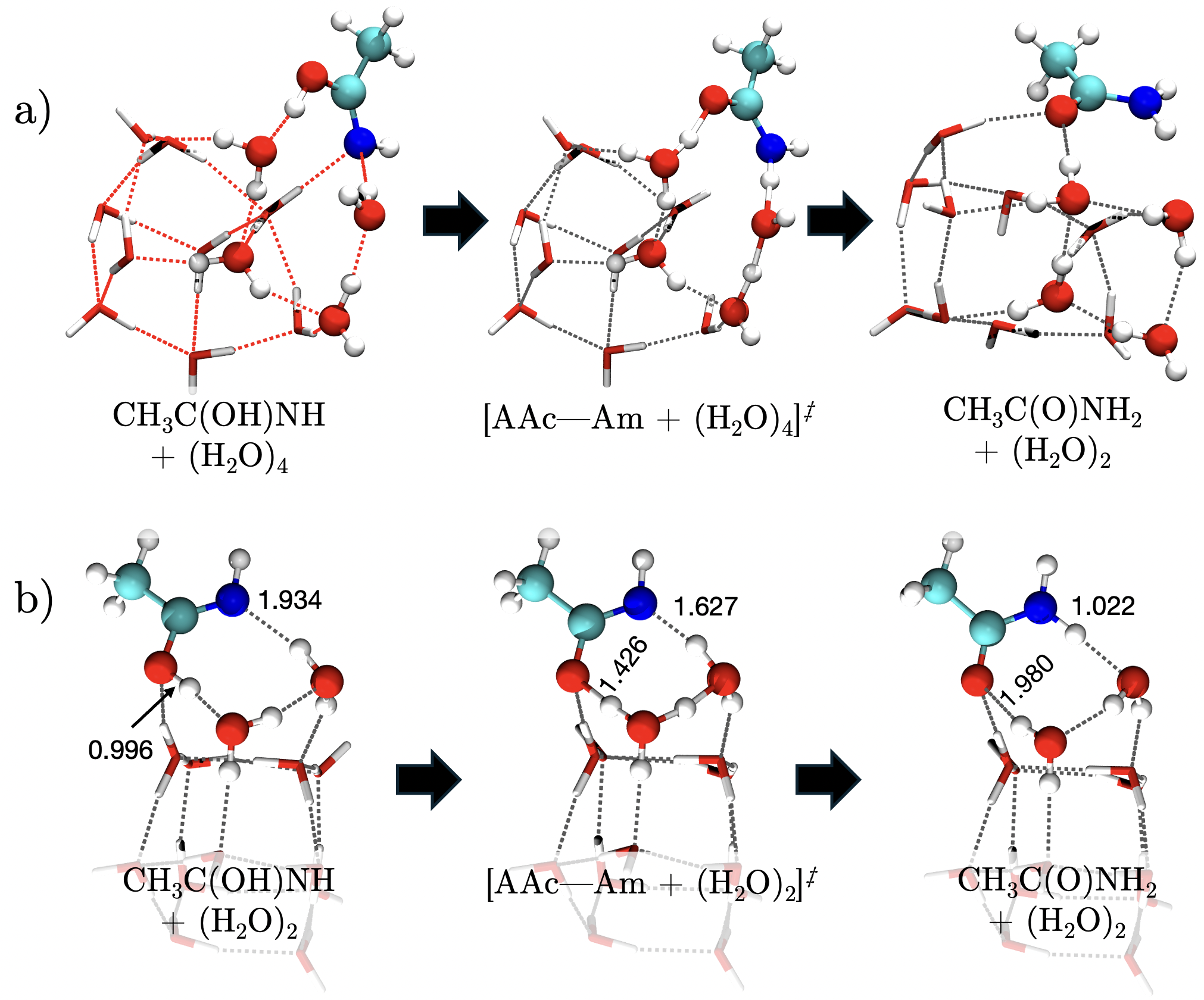}
    \caption{Snapshots of the potential energy surface stationary points for the acetimidic acid to acetamide isomerisation. The upper panel, a), corresponds to the high barrier isomerisation reaction case, while the lower panel, b), to the lower barrier one. Distances are provided in \AA. Atoms relevant to the reaction are highlighted using a ball-and-stick representation. Atom colour code is: red for O, blue for N, grey for C and white for H.}
    \label{fig:AAc_Am}
\end{figure}

Finally, an alternative reaction path involves the direct H-transfer from the intermediate \ce{OH-^.C=NH} radical by \ce{^.CH3}. Such a reaction would lead to either cyanic acid (\ce{HOCN}) or isocyanic acid (\ce{HNCO}), see Figures \ref{fig:wht_4water}(b) and \ref{fig:cyanic_acid}. The barriers for each of these channels are 23.3 and 33.2~\kjmol, respectively, indicating that the coupling reaction leading to \ce{CH3C(OH)NH}) is far more efficient. For the sake of comparison, and given the potential importance of cyanic and isocyanic acid on interstellar ices, whose conjugate base is \ce{OCN-}, commonly detected in interstellar ice IR adsorption bands \citep[e.g.,][]{Boogert2015,McClure2023} we have also studied these reactions substituting \ce{^.CH3} by an H atom, which results in the lowering of the barriers down to 6.0 and 23.2~\kjmol, for the formation of cyanic and isocyanic acid, respectively (see Figure \ref{sifig:HOCNH+H} in the appendix). The barriers for each reaction keep the same relative ordering as for the reactions with the methyl radical. Still, again, this mechanism would have to compete with the H-addition leading to 
\ce{HOCHNH} (hydroxymethylimine). Therefore, these seem to be rather inefficient ways to form cyanic and isocyanic acid.
Coming back to acetamide, our proposed mechanism aligns with the pathway suggested by \citet{Belloche2017_acetamide}, linking acetamide and formamide, and also incorporates elements of the ice-surface processing route described by \citet{Drabkin2023MNRAS}, namely, we propose acetimidic acid as a precursor of acetamide.
Once acetimidic acid is formed, it can undergo keto-enol tautomerism facilitated by the ice surface, converting it into acetamide through a water-assisted H-transfer reaction. To explore this, we manually prepared two binding modes shown in Figure \ref{fig:AAc_Am}a) and b) in a way that there is a clear H-transfer relay path. The differences between the two modes are the interaction of AAc with the cluster and the number of water molecules participating in the isomerisation. The first binding mode exhibits a high activation barrier of 58.1 \kjmol and a low crossover temperature of 58~K, with a reaction occurring over four water molecules. In contrast, the second binding mode presents a significantly lower barrier of just 20.0 \kjmol, comparable to the lowest energy barrier for the \ce{(^.CN-H2O)_{hemi} \to HO^.CNH} conversion, a crossover temperature of 134~K and requires only two water molecules. This highlights the critical role of the binding mode and water's catalytic influence in facilitating H-transfer reactions and the strong impact of local ice structure on reaction energetics. 

\myhlight{On the other hand}, on CO ices the picture is much simpler. The reaction of \ce{^.CH3} with either \ce{^.CN} and \ce{NC^.CO} (the product of cyanide's chemisorption on CO ice) is barrierless, leading to \ce{CH3CN}, \ce{CH3NC} or \ce{CH3C(O)CN} (acetyl cyanide). In these cases, competition between diffusion and reaction is also expected, especially given the much-reduced binding energies on CO ices. For example, \ce{^.CH3} has a binding energy of 1.7~\kjmol ($\sim$200~K) on CO \citep{Lamberts2019}, while it takes values between 8.3--13.3~\kjmol ($\sim$1000--1600~K) on amorphous solid water \cite{Ferrero2020,Bovolenta2022}. 

\myhlight{Finally, we would like to highlight that, despite the technical challenges associated with working with nitriles (e.g., HCN) in the laboratory (mainly due to their toxicity) and the inherent difficulties of handling radical species, exploring these reaction pathways experimentally would provide valuable insights into the final products of the overall reaction network. To date, no dedicated experimental studies have addressed the formation of \ce{CH3CN} and \ce{CH3NC} on interstellar ice analogues other than those focused on the energetic processing of ices \citep[e.g.,][]{Volosatova2021,Canta2023,Chuang2024}. Regardless, the formation of \ce{CH3CN} on interstellar ices remains a particularly intriguing subject, as this species is believed to serve as a precursor to more complex molecules, such as amino acids, amines, and amides \citep[e.g.,][]{Hudson2008,Danger2011_VUV_CH3CN,Bulak2021}. As such, further experimental investigation is certainly warranted.}

\section{Conclusions}\label{sec:conclusions}

This work further supports the idea that depleted \ce{^.CN} radicals on interstellar dust grains cannot be "preserved" for future chemical processes if they remain on the surface. In \citet{enrique2024complex}, we demonstrated that \ce{^.CN} is readily converted into HCN or HNC, except in environments with insufficient atomic H. Here, we show that \ce{^.CN} is also highly reactive with fast-diffusing radicals like \ce{^.CH3}, as \ce{^.CN} itself is expected to be largely immobile due to its unusually strong hemibonded interaction with water ice surfaces. The reaction \ce{^.CH3 + ^.CN} leads to the formation of both methyl cyanide (\ce{CH3CN}) and methyl isocyanide (\ce{CH3NC}), with these reactions competing against \ce{^.CH3} back-diffusion. Additionally, even if \ce{^.CN} does not react to form HCN, HNC, \ce{CH3CN}, or \ce{CH3NC} on water ices, it can still react with a water molecule to form \ce{HO^.CNH}, as shown earlier by \citet{Rimola2018}. This newly formed radical is related to formamide (\ce{HCONH2}) via a reaction with H, acetimidic acid (\ce{CH3COHNH}) via reaction with \ce{CH3}, and acetamide (\ce{CH3CONH2}) following a water-assisted isomerisation.

On the other hand, on CO ices, both \ce{CH3CN} and \ce{CH3NC} form without an energy barrier. Likewise, \ce{CH3C(O)CN} can form when \ce{^.CN} reacts with surface CO, leading to \ce{NC^.CO}. As with previous cases, these reactions compete with radical diffusion.

Finally, we want to point out that there are more surface routes to form methyl cyanide, such as \ce{^.CCN} hydrogenation, which will be the focus of a forthcoming work, \myhlight{and that further experimental studies on these reaction paths would certainly be of interest to the community.}

\paragraph{Data availability} Extra supporting information can be found in the Zenodo repository with DOI: 10.5281/zenodo.15630762 (\href{https://doi.org/10.5281/zenodo.15630762}{link}).

\begin{acknowledgement}
We thank Dr. G. Molpeceres for stimulating discussions, especially regarding the AVAS calculations in {\sc Molpro}.
This project has received funding from the Horizon Europe Framework Programme (HORIZON) under the Marie Skłodowska-Curie grant agreement No 101149067, ``ICE-CN''. 
Finally, this work was granted access to the HPC resources of the high-performance computer SNELLIUS, part of the SURF cooperative of educational and research institutions in The Netherlands, under project No EINF-6197.
\end{acknowledgement}

\bibliographystyle{aa}
\bibliography{mybiblio}

\begin{appendix}

\section{Benchmark}\label{sisec:benchmark}

This section presents the structures used in our benchmark study; corresponding energetic data are available in the Zenodo repository (\href{https://doi.org/10.5281/zenodo.15630762}{link}).
For multireference calculations, active space selection varied by software. In {\sc Molpro}, we used an automated method to include relevant atomic orbitals: the 2p$_z$ of C in \ce{CH3}, the 2p$_i$ of O in water aligned with CN hemibonding, and the full valence of CN (alignment ensured proper selection). A figure showing the resulting orbitals before and after the CASPT2 step is available from the online repository, named as Figure A3. 
In {\sc OpenMolcas}, full-valence selection was limited by memory, so we manually defined an active space of 24 orbitals and 15 electrons from canonical UHF orbitals, focusing on orbitals centred on reacting atoms.

\begin{figure}[!htbp]
    \centering
    \includegraphics[width=0.8\columnwidth]{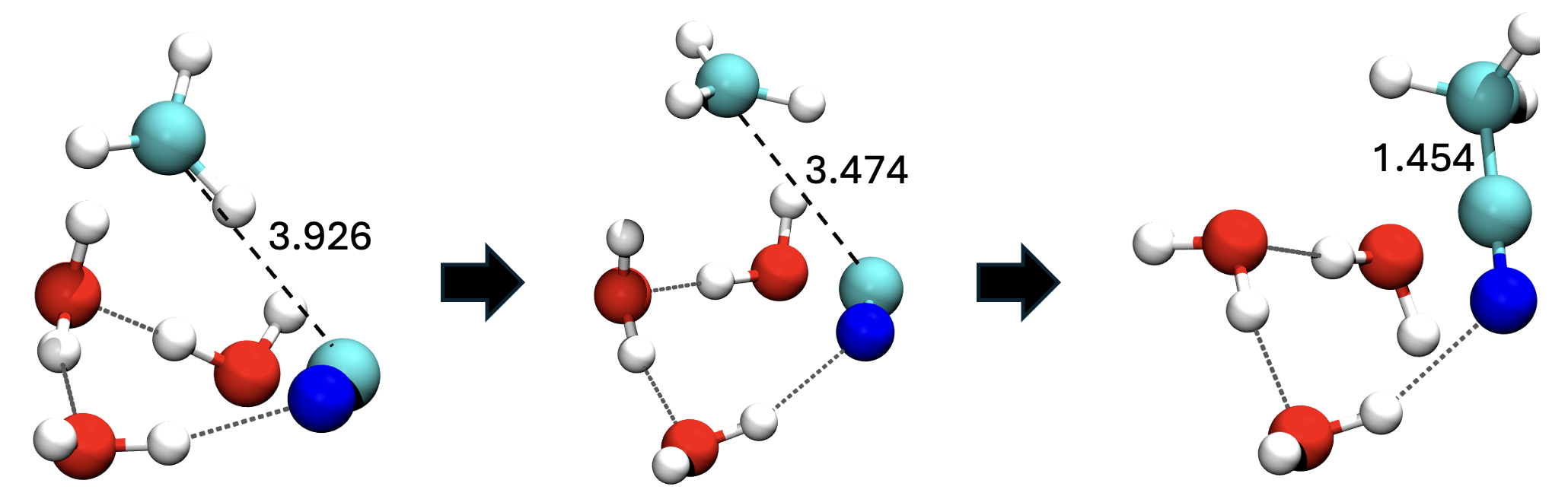}\\
    {\footnotesize \textbf{a)} \ce{CH3CN} formation on a water trimer.\\}
    \vspace{0.7cm}
    
    \includegraphics[width=0.8\columnwidth]{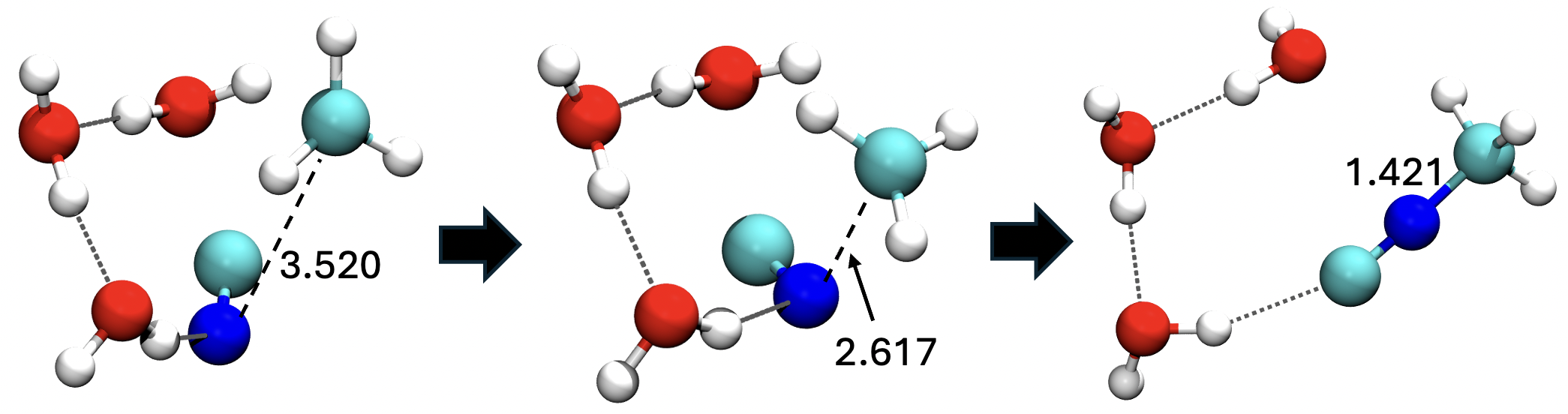}\\
    {\footnotesize \textbf{b)} \ce{CH3NC} formation on a water trimer.\\}
    \vspace{0.7cm}
    
    \includegraphics[width=0.9\columnwidth]{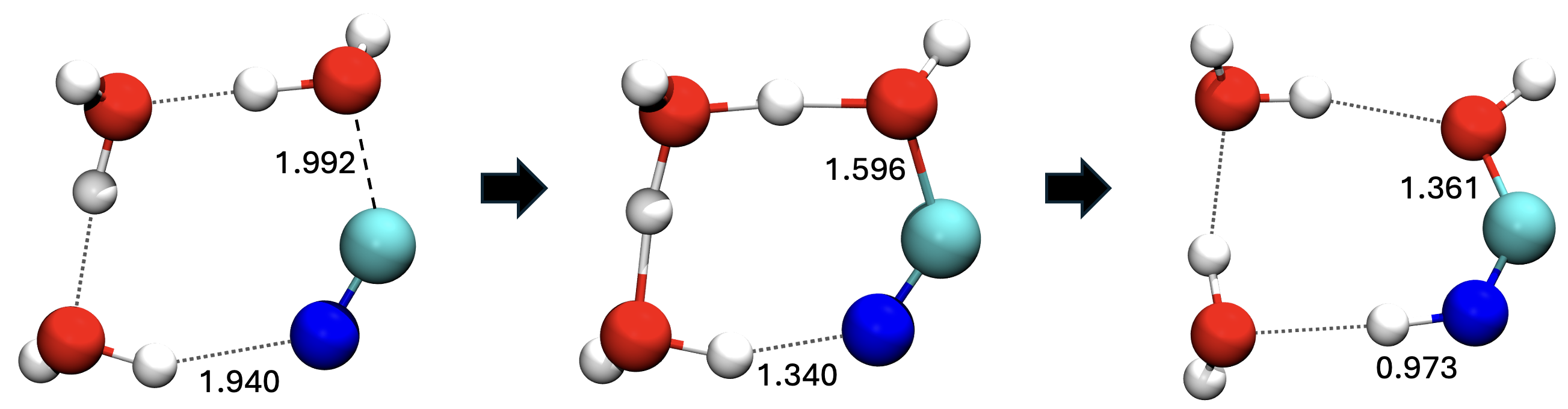}\\
    {\footnotesize \textbf{c)} water-assisted H-transfer to form \ce{HO^.C=NH}.\\}
    \caption{Structures used in the reactivity benchmark. Atomic colouring code: white for hydrogen, grey for carbon, blue for nitrogen and red for oxygen.}
    \label{sifig:enter-label}
\end{figure}

\begin{figure}[!htbp]
    \centering
    \includegraphics[width=\columnwidth]{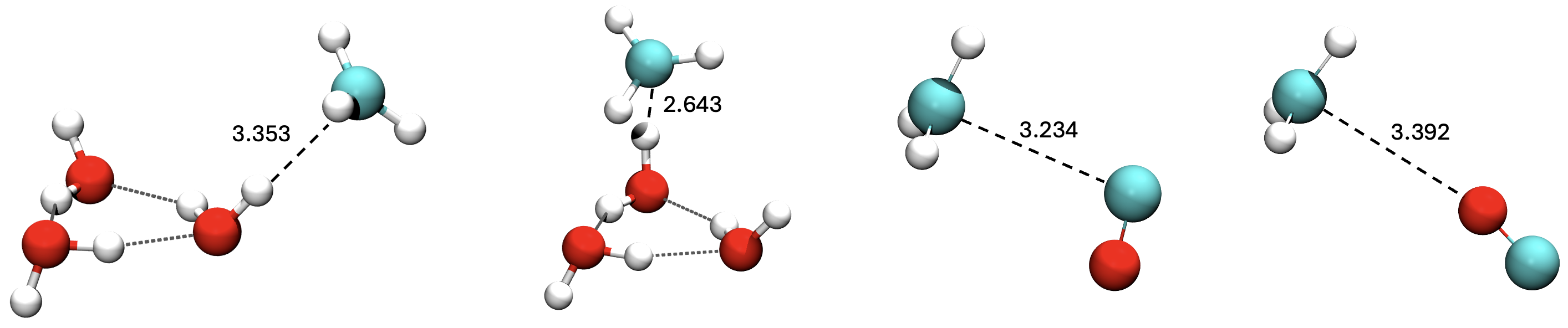}
    \caption{Structures used in the interactions benchmark. From left to right: cases 1 and 2 on the water trimer, and cases 1 and 2 on a single CO molecule. Atomic colouring code: white for hydrogen, grey for carbon and red for oxygen.} 
    \label{sifig:enter-label}
\end{figure}

\section{Methyl radical diffusion}

\begin{figure}[H]
    \centering
    \includegraphics[width=0.49\columnwidth]{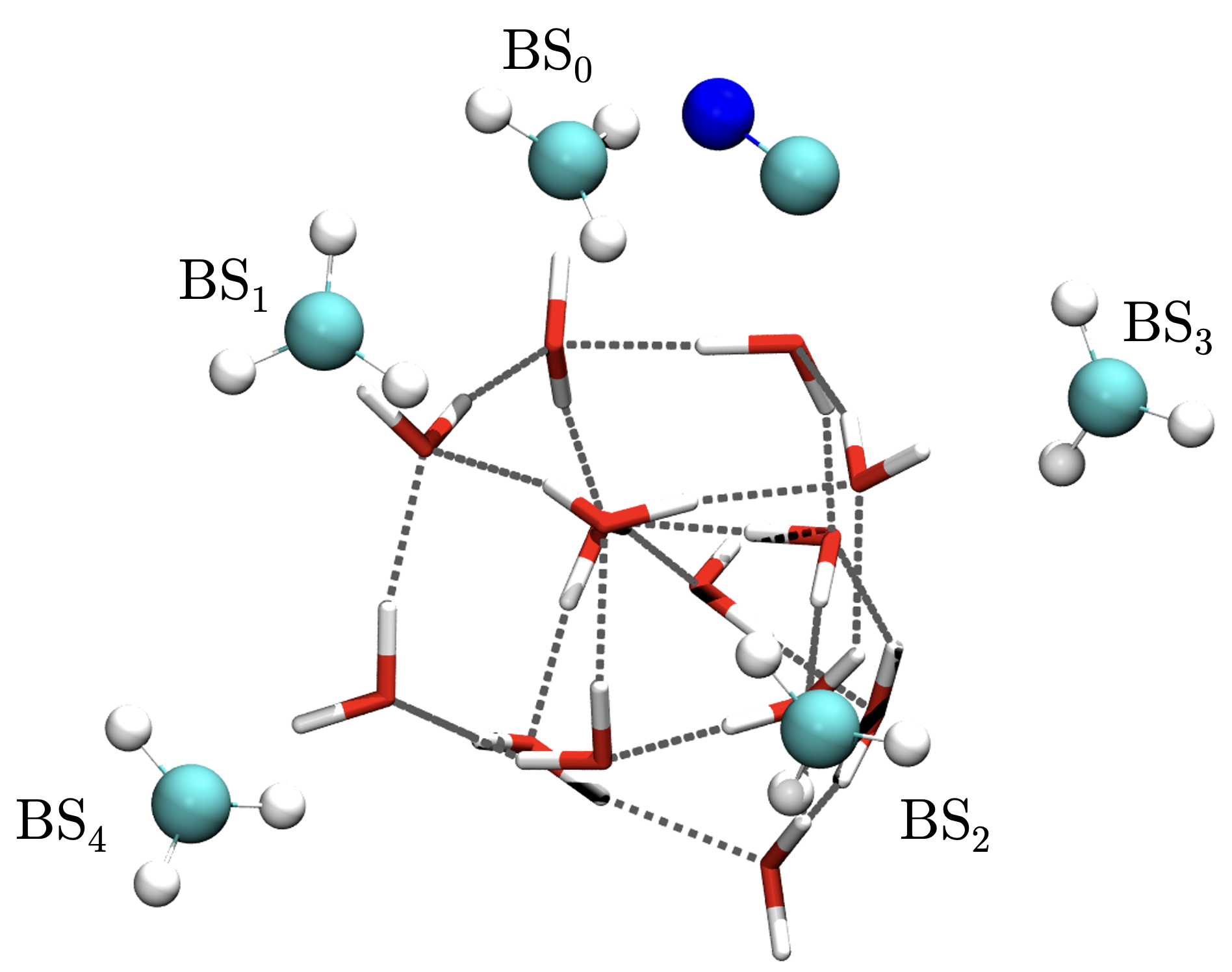}
    \includegraphics[width=0.49\columnwidth]{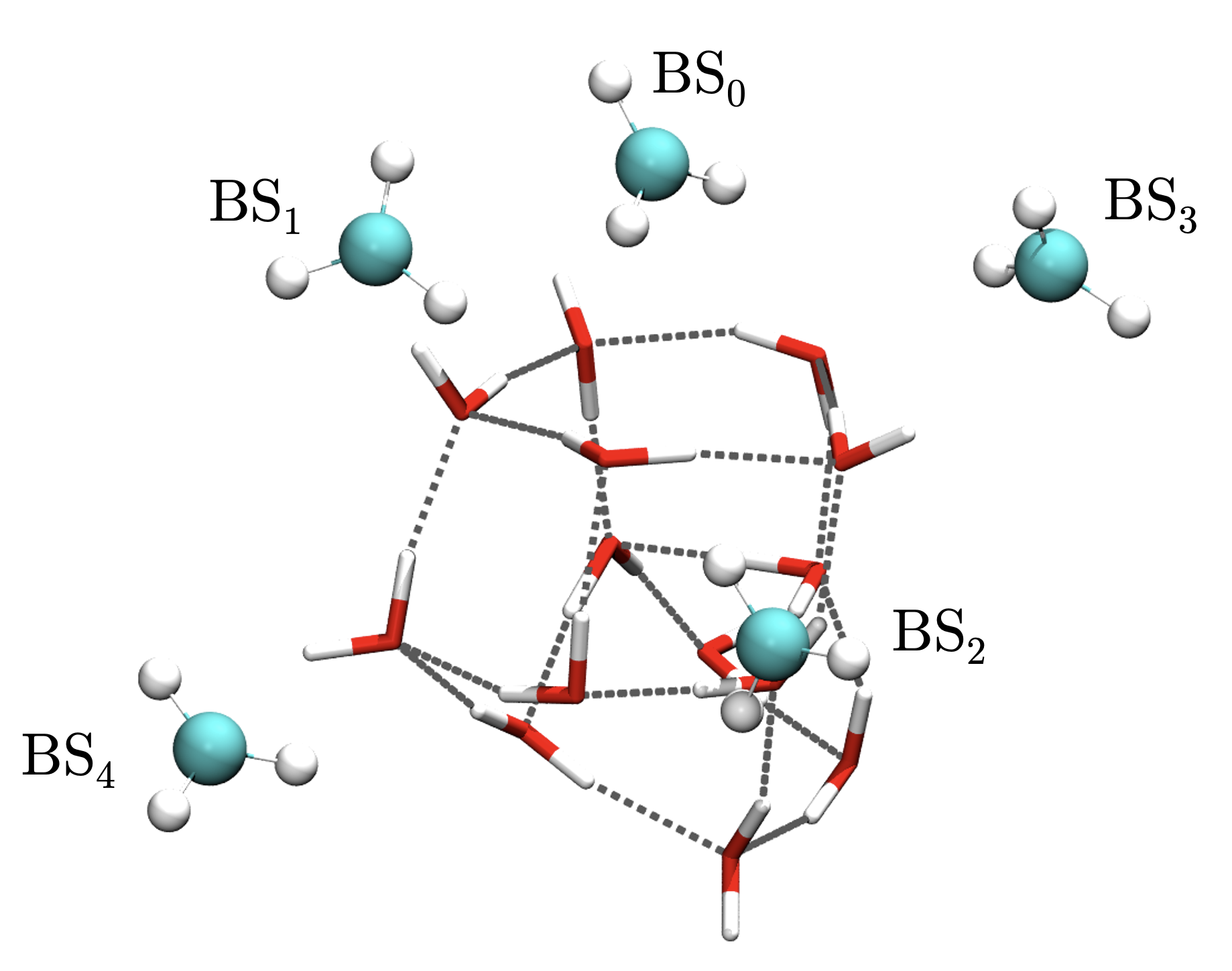}
    \caption{Binding sites explored to study the diffusion of \ce{^.CH3} on water with (left panel) and without (right panel) CN on the surface. Atomic colouring code: white for hydrogen, grey for carbon, blue for nitrogen and red for oxygen.}
    \label{sifig:SIdiff}
\end{figure}

Figure~\ref{sifig:SIdiff} shows the binding sites used in the diffusion study. Energetic and transition state data are provided in Table B1 in the Zenodo repository (\href{https://doi.org/10.5281/zenodo.15630762}{link}).

\section{Figures for \ce{HO^.CNH + ^.H\to HOCN/HNCO + H2}}

\begin{figure}[H]
    \centering
    \includegraphics[width=\columnwidth]{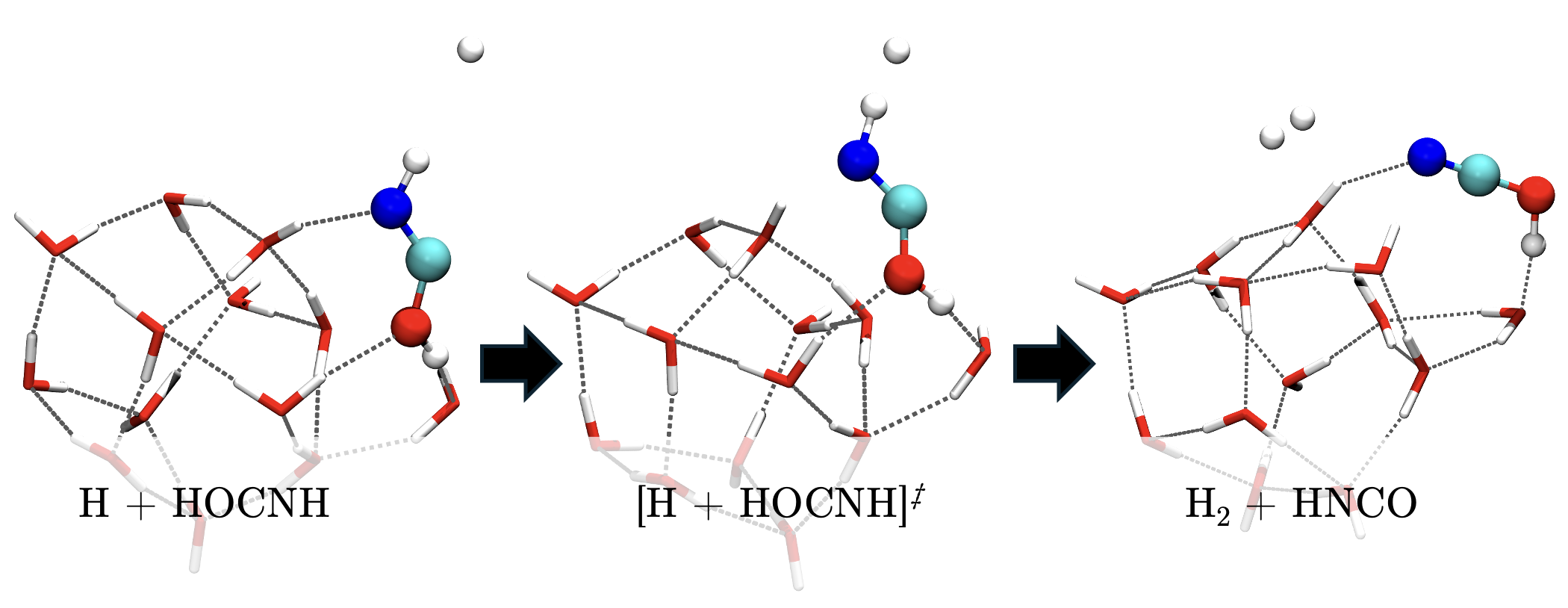}\\
        \textbf{(a)} \ce{HO^.CNH + ^.H\to HOCN + H2}\\
    \includegraphics[width=\columnwidth]{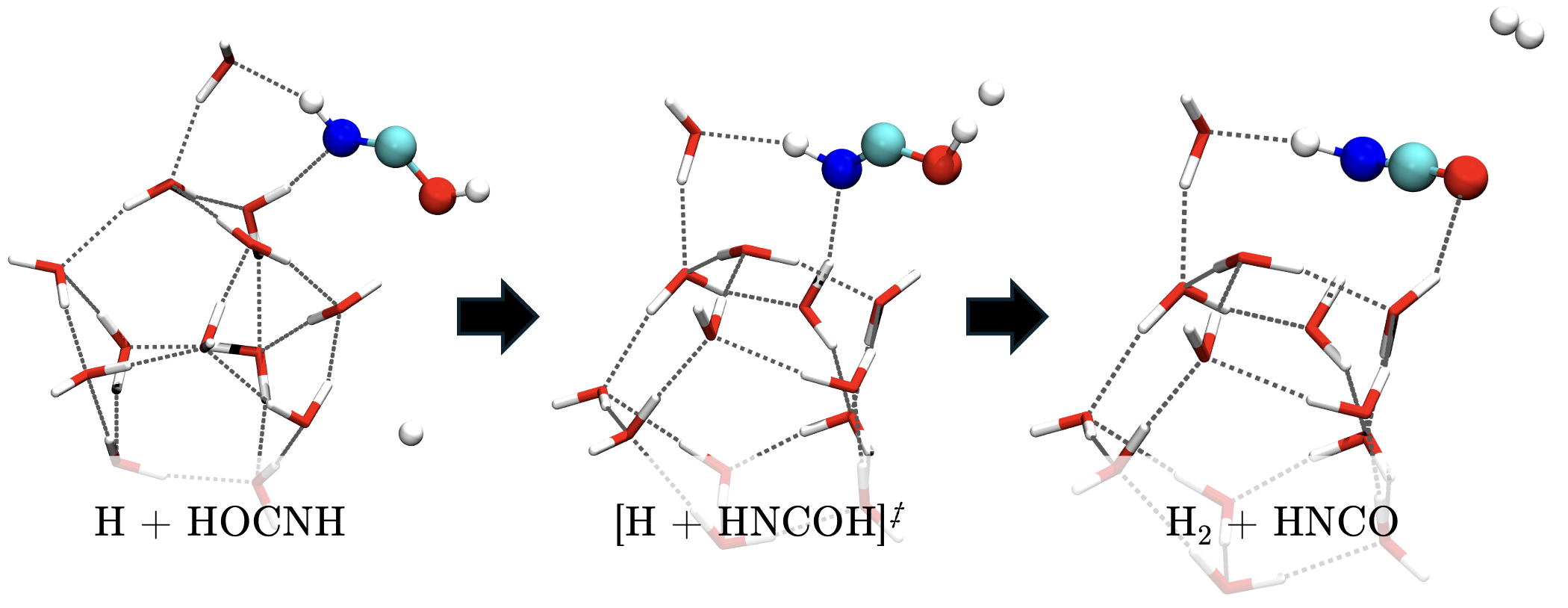}\\
        \textbf{(b)} \ce{HO^.CNH + ^.H\to HNCO + H2}\\
    \caption{Steps of the H-abstraction reactions to form cyanic and isocyanic acid from \ce{HO^.CNH + ^.H}. Atomic colouring code: white for hydrogen, grey for carbon, blue for nitrogen and red for oxygen. The energetics can be found in the repository (\href{https://doi.org/10.5281/zenodo.15630762}{link}), in Table D4.}
    \label{sifig:HOCNH+H}
\end{figure}

We have studied the H-abstraction by \ce{^.H} from \ce{HO^.CNH} by just taking the structures for the reactions with \ce{^.CH3} and substituting the methyl radical by the hydrogen atom, then finding the stationary points of the reaction, as shown in Figure \ref{sifig:HOCNH+H}.

\section{Reactivity with molecules from the ice} \label{sisec:energetics_onice}

Tables D1–D4 in the Zenodo repository (\href{https://doi.org/10.5281/zenodo.15630762}{link}) report activation and reaction energies. Table D1 covers radical–radical couplings to form \ce{CH3CN} and \ce{CH3NC}, while Tables D2–D4 include water-assisted H-transfers and their follow-up reactions: D2 (wHt over 2–3 waters), D3 (wHt over 4 waters + abstraction by \ce{^.CH3}), and D4 (abstraction by H atoms).

\end{appendix}

\end{document}